\documentstyle[11pt,newpasp,twoside,epsf]{article}
\def\edcomment#1{\iffalse\marginpar{\raggedright\sl#1\/}\else\relax\fi}
\reversemarginpar

\begin{document}
\title{The Frequency Content of the VIRGO/SoHO Lightcurves: 
       Implications for Planetary Transit Detection from Space}
\author{S.~Aigrain$^{1}$, G.~Gilmore$^{1}$, F.~Favata$^{2}$, S.~Carpano$^{2}$}
\affil{$^{1}$Institute of Astronomy, University of Cambridge, Madingley Road, 
             Cambridge, CB3 0HA, United Kingdom}
\affil{$^{2}$Astrophysics Division, Space Science Department, ESTEC, 
             P.O.~Box 299, Noordwijk, 2200 AG, The Netherlands}

\begin{abstract}
Stellar micro-variability poses a serious threat to the capacities of 
space-based planet-finding missions such as \emph{Kepler} or \emph{Eddington}. 
The solar lightcurves obtained by the VIRGO PMO6 and SPM instruments on board 
SoHO from 1996 to 2001 have been studied in order to follow variability changes
through the activity cycle. In all datasets, active regions-induced 
variability, below $2~\mu\mathrm{Hz}$, is closely correlated to the BBSO 
Ca~{\sc II} K-line index. 
The PMO6 (total irradiance) data shows evidence for a meso-granulation 
component around $\tau \simeq 8 \times 10^3~\mathrm{s}$, while all narrow-band 
SPM datasets (red, green and blue) show super-granulation ($\tau \simeq 5 
\times 10^4 ~\mathrm{s}$) but no meso-granulation. Both actvity and 
granulation related components have 
significantly smaller amplitudes in the red than in the blue channel. 
These results, coupled with available stellar data, allow us to generate 
simulated lightcurves with enhanced variability as a testbed for pre-processing
and detection methods, and influence the case for using colour information in 
this kind of mission.
\end{abstract}

\section{Introduction}

Transit detection algorithms in photometric time series, developed by the 
COROT (Defa{\" y}, Deleuil, \& Barge 2001), \emph{Eddington} 
(Aigrain \& Favata 2002) and \emph{Kepler} (Jenkins, Caldwell, \& Borucki 2002)
teams are very effective in the presence of white noise, and show that the 
latter two missions could detect habitable Earth-sized planets in data with
white noise only. 
However, the non-Gaussian, ill-known stellar variability can have amplitudes 
over an order of magnitude larger than an Earth-analogue transit. Simulations 
using PMO6 data and the Bayesian algorithm developed for \emph{Eddington} show
that the minimum detectable planet size changes to $3~R_{\oplus}$ when Sun-like
variability is introduced without sophisitcated filtering 
techniques\footnote{Simulations for a K5V star with $V=14.5$, 4 
transits in the light curve} (Aigrain, Gilmore, \& Favata, 2001). Variability 
will thus impact the choice of pre-processing techniques as well as target 
selection and colour information issues. 

Pre-processing methods are already under investigation. After testing a 
highpass filter with some success, the \emph{Eddington} team have developed a 
more optimized filter, based on a priori knowledge of the transit shape. 
Initial tests suggest it will be very successful, but realistic lightcurves 
with various levels of activity are needed for a quantitative assessment. The 
only star monitored with sufficient photometric precision and sampling is the 
Sun, which we use as a starting point.

\section{Method}
 
The power spectrum of the Sun's lightcurve contains, beside the sharp 
oscillatory peaks used in helioseismology, an underlying `solar background'. 
It is mostly concentrated at low frequencies and has a complex, 
multi-component structure extending to a few mHz. It is commonly modeled as a 
sum of power laws (Harvey 1985, Andersen 1991, Andersen 1992, 
Harvey et al.~1993):
\begin{displaymath}
P(\nu) = \displaystyle{\sum_{i=1}^{N}} P_{i}(\nu) = 
         \displaystyle{\sum_{i=1}^{N}} \frac{\mathrm{A}_{i}}
                                     {1 + (\mathrm{B}_i \times \nu)^{\mathrm{C}_i}}
\end{displaymath}
Various authors (Harvey et al.~1993, Andersen, Leifsen, \& Toutain 1994, 
Rabello Soares et al.~1997) found up to 5 components in total irradiance:
active regions, with characteristic timescale $\tau = 1$ to $3 \times 10^5 
~\mathrm{s}$; super-granulation ($\tau = 3$ to $7 \times 10^4~\mathrm{s}$); 
meso-granulation ($\tau \simeq 8000~\mathrm{s}$); granulation 
($\tau = 200 to 500~\mathrm{s}$); bright points ($\tau \simeq 70~\mathrm{s}$).
Interestingly, analysis of the narrow channel SPM data to date has shown no 
evidence for a meso-granulation component (Andersen et al.~1998, Pall{\' e}, 
Roca Cort{\' e}s, \& Jim{\' e}nez 1999). 

With the VIRGO dataset now spanning the interval from solar minimum to maximum,
the evolution of the various components with the activity cycle can be studied.
An algorithm has been developed to fit the Fourier Transform (FT) of lightcurve
sections of duration $L$, adding components one by one until adding a component
no longer improves the fit. A new section, offset from the previous one by a 
step $S$, is then fitted in turn, using the last fit as an initial guess. An 
example fit is shown in Fig.~1(a). Thus we monitor variations in amplitude, 
timescale and slope for each component and attempt to relate them to 
observables measured in many stars.

\begin{figure}
\plotone{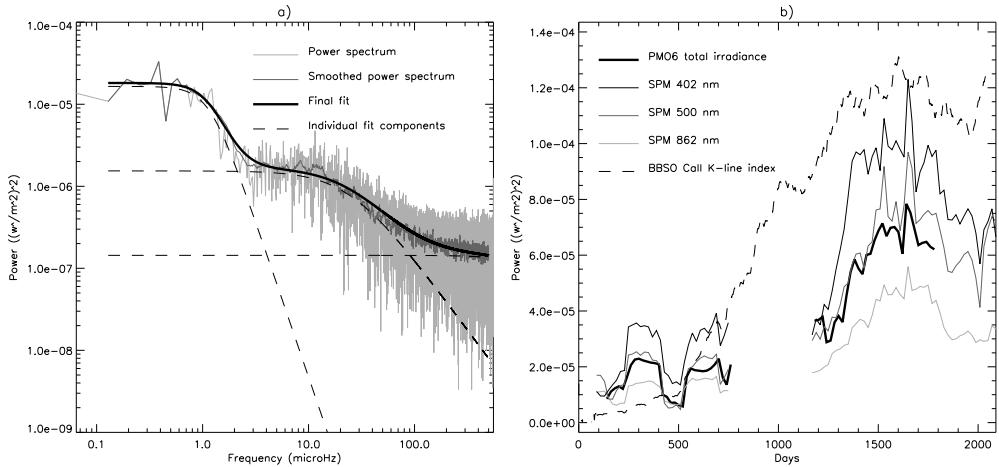}
\caption{(a) Example fit to a 180 day section of PMO6 data. 
(b) Evolution of the amplitude of the active regions component between 
1996 and 2001, using $L = 180$ days, $S = 20$ days. The BBSO Ca~{\sc II} K-line 
index has been smoothed over 180 days, scaled down and offset for clarity.
The gap at around 1000 days correspond to a prolonged gap in the data. }
\end{figure}

\section{Results}

The algorithm described above was run on the PMO6 and SPM data with $L = 180$ 
days and $S = 20$ days, resulting in two components plus a third whose turnoff 
was unresolved at higher frequencies. A number of points of interest emerge 
from the results. In all 4 datasets the amplitude of the first component, with 
$\tau \simeq 1.3 \times 10^{5}~\mathrm{s}$ (active regions) increases from 
solar minimum to maximum and is correlated with the Ca~{\sc II} K-line index, 
indicator of chromospheric activity (see Fig.~1(b)). The amplitude of the 
second component also increases, but is not correlated to the Ca~{\sc II} 
index.  In both cases, the red channel shows a lower amplitude than the blue 
and green channels. 
Noticeably, the timescales are different in the PMO6 and the SPM data. The 
difference is significant for the second, granulation related component, with 
$\tau \simeq 8 \times 10^3~\mathrm{s}$ (meso-granulation) and $\tau \simeq 5 
\times 10^4~\mathrm{s}$ (super-granulation) for PMO6 and SPM data respectively.
This may result from a difference in contrast between super- and 
meso-granulation in the different channels, as suggested by Andersen et 
al.~(1998). The meso-granulation signal present in total irradiance may also 
be due to spectral features not included within the narrow SPM channels. This 
question will be investigated by examining the spectral content of the 
channels and comparing solar disk images to the components' evolution. The 
slopes are 3.8 (active regions) and 1.75 (granulation, both datasets), in good 
agreement with  Andersen et al.~(1998).

\section{Discussion and future work}

We have shown that indicators of chromospheric activity such as the 
Ca~{\sc II} K-line index provide a good proxy to the amplitude of the Sun's 
variability at low frequencies. As similar indicators are commonly measured in 
other stars, we may infer their expected variability level up to 
a few $\mu\mathrm{Hz}$. However, little theoretical or observational insight is 
yet available to relate granulation-type phenomena to stellar observables.
Relating the observed components in the power spectrum to spectral and/or 
surface features may allow us to identify some new indicators of this type of 
variability. 

In the mean time, a lightcurve `simulator' is being developed, using known 
scaling laws between chromospheric indicators, rotation period and colour, in 
which we can realistically scale up the active regions component. In the 
absence of more information a variety of scaling laws will be tried for the 
granulation component. This will provide a thorough test of pre-processing 
techniques under development.

In the longer term missions such as MOST, MONS and COROT will provide an ideal 
dataset to calibrate a general micro-variability model\footnote{A proposal for 
the COROT Parallel Science Working Group, entitled \emph{Hours-timescale 
stellar background variability and exo-planet transit detection prospects} 
(PI A. Collier-Cameron) has recently been accepted.}. Our understanding of the 
mechanisms behind low frequency, low amplitude stellar variability promises to 
improve vastly over the next few years.
        
Work is underway to quantify any increase in detection power gained from 
concentrating on the red part of the spectrum where we observe smaller 
variability, and from optimizing the filter bandwidth to exclude particularly
variable spectral regions. These must be weighed against loss of 
photometric accuracy.

\acknowledgements The authors wish to thank the staff at Big Bear Solar 
Observatory for kindly providing us with their Ca~{\sc II} K-line index data, 
and the VIRGO Science team for access to the PMO6 and SPM data. This work was 
supported in part by the European Space Agency's Young Graduate Trainee 
program and by studentships from the UK's Particle Physics and Astronomy 
Research Council and the Isaac Newton Trust.


\begin{references}

\reference Aigrain, S., Gilmore, G., \& Favata, F.~2001, in Techniques for the 
           Detection of Planets and Life beyond the Solar System, 4th Annual 
           ROE Workshop ed.~W.~R.~F.~Dent, 8
\reference Aigrain, S., \& Favata, F.~2002, \aap, submitted
\reference Andersen, B.~N.~1991, Adv.~Sp.~Res., 11, 4, 93
\reference Andersen, B.~N.~1992, in Proceedings of Mini-Workshop on Diagnostics
           of Solar Oscillations Observations (University of Oslo), 15
\reference Andersen, B.~N., Leifsen, T., Toutain, T.~1994, So.~Ph., 152, 247
\reference Andersen, B.~N., Appourchaux, T., Crommelnynck, D., Fr{\" o}hlich, 
           C., Jim{\' e}nez, A., Rabello Soares, M.~C., \& Wehri, Ch.~1998, 
           IAU Symp.~181
\reference Defa{\" y}, C., Deleuil, M., \& Barge, P.~2001, \aap, 365, 330
\reference Harvey, J.~W.~1985, ESA SP-235, 199
\reference Harvey, J.~W., Duvall, T.~L., Jefferies, S.~M., \& Pomerantz, M.~A.~
           1993, in ASP Conf.~Ser., 42, 111
\reference Jenkins, J.~M., Caldwell, D.~A., \& Borucki, W.~J.~2002, \apj, 564, 
           495
\reference Pall{\' e}, P.~L., Roca Cort{\' e}s, T., Jim{\' e}nez, A., \& GOLF 
           \& VIRGO Teams 1999, ASP Conf.~Ser., 173, 297
\reference Rabello Soares, M.~C., Roca Cort{\' e}s, T., Jim{\' e}nez, A., 
           Andersen, B.~N., \& Appourchaux, T.~1997, \aap, 318, 970

\end{references}
\end{document}